\documentclass[manuscript, nonacm]{acmart}

\setcopyright{none}
\renewcommand\footnotetextcopyrightpermission[1]{}

\AtBeginDocument{%
  }

\usepackage{tikz}
\usepackage{xcolor}
\usepackage{graphicx}

\usetikzlibrary{
    positioning,
    arrows.meta,
    calc,
    backgrounds
}

\makeatletter
\def\@authorsaddresses{}
\makeatother
\begin{document}

\title{It Was Better Not to Know: Digital Recognition, Family Authority, and Postpartum Mental Health in Bangladesh}

\author{Shajnin Haque Aishy}
\affiliation{%
  \institution{Computing for Sustainability and Social Good (C2SG) Research Group, United International University}
  \city{Dhaka}
  \country{Bangladesh}
}
\affiliation{%
  \institution{Department of Computer Science and Engineering, BRAC University}
  \city{Dhaka}
  \country{Bangladesh}
}
\email{shajnin.haque.aishy@g.bracu.ac.bd}

\author{Sabbir Bin Abdul Latif}
\affiliation{%
  \institution{Computing for Sustainability and Social Good (C2SG) Research Group, United International University}
  \city{Dhaka}
  \country{Bangladesh}
}
\affiliation{%
  \institution{Institute of Information Technology, University of Dhaka}
  \city{Dhaka}
  \country{Bangladesh}
}
\email{msse2011@iit.du.ac.bd}

\author{Dr. Md. Golam Zel A Husna}
\affiliation{%
  \institution{National Institute of Cancer Research and Hospital}
  \city{Dhaka}
  \country{Bangladesh}
}

\author{Jannatun Noor}
\affiliation{%
  \institution{Computing for Sustainability and Social Good (C2SG) Research Group, Department of Computer Science and Engineering, United International University}
  \city{Dhaka}
  \country{Bangladesh}
}
\email{jannatun@cse.uiu.ac.bd}

\renewcommand{\shortauthors}{Shajnin et al.}
\renewcommand{\authorsaddresses}{}

\begin{abstract}
  Postpartum mental distress affects roughly one in five mothers in low- and lower-middle-income countries, yet recognition and support remain limited where formal mental-health care is scarce, and family relationships govern health decisions. We report a five-month interview study with 20 mothers in urban, urban slum, and rural Bangladesh who had given birth within the previous two years. Mothers encountered the concept of postpartum depression incidentally, through Facebook and YouTube recommendations rather than deliberate search or clinical guidance, and then used these platforms for private sensemaking without disclosing. Recognition, however, rarely traveled beyond the self: mothers could identify their distress while families dismissed it, leaving them with knowledge but no authority to act. Participants sought validation as a precondition for support and described recovery as relational rather than clinical or digital. We argue that postpartum technologies should engage partners, families, and community health workers rather than placing recognition solely on mothers.
\end{abstract}





\maketitle

\section{Introduction} \label{sec:introduction}
A mother in a typical Bangladeshi household scrolls through Facebook when a video names something she has felt for months but had no word for. Crying without a clear reason, numbness toward her own baby, a wish to disappear from her own life. She tells her family, but her mother-in-law dismisses it as ordinary motherhood, and her husband questions what she read online. She stops talking. She now understands her distress, but nothing else has changed. This gap—between a mother’s recognition of her distress and her household’s willingness to recognize it—is the subject of this paper.
 
Postpartum mental distress covers the emotional and psychological difficulties that can follow childbirth \cite{ohara2013postpartum, perry2021phenomenology}, from short-lived sadness and exhaustion to more persistent conditions such as postpartum depression \cite{perry2021phenomenology, beck2001predictors}. Left unaddressed, it can affect a mother's daily life, relationships, caregiving, and long-term wellbeing \cite{slomian2019consequences, lovejoy2000maternal}, yet many mothers do not recognize these experiences as a mental-health concern or seek support for them \cite{dennis2006postpartum, perinatalmhl2022}. The scale is large: roughly one in five women in low- and lower-middle-income countries experiences a common perinatal mental disorder, substantially higher than in high-income countries \cite{fisher2012prevalence, wang2021mapping}, with untreated perinatal depression linked to higher maternal mortality \cite{hagatulah2024perinatal}, worse infant development \cite{stein2014effects, kingston2012prenatal}, and economic costs for families \cite{bauer2024economic}. Stigma, normalization, and uncertainty about when distress needs support all delay recognition and help-seeking \cite{dennis2006postpartum, perinatalmhl2022}. We use postpartum mental distress broadly here, to include experiences that may or may not receive a clinical diagnosis of postpartum depression.
 
Clinical psychology, psychiatry, nursing, and global mental health have studied postpartum mental health for decades \cite{beck2001predictors, ohara2013postpartum, fisher2012prevalence}, showing that family involvement matters for recovery, particularly in South Asia \cite{rahman2008cognitive, sikander2019delivering, cluxtonkeller2018clinical}. HCI has approached postpartum mental health differently, mostly through computational detection \cite{dechoudhury2014characterizing, king2019microstress, maharjan2021passive}, online disclosure and communities \cite{progga2023understanding, andalibi2020disclosure, ammari2019throwaway}, and digital interventions such as apps and conversational agents \cite{barry2017mhealth, milgrom2016internet, suharwardy2020effect}. This study has taught us a great deal about reaching, screening, and supporting individual mothers online, but it centers the individual mother and was conducted in Western or relatively well-resourced settings \cite{linxen2021weird}, saying less about settings where formal care is scarce and family relationships govern access to healthcare, technology, and health decisions \cite{sultana2018design, mustafa2020patriarchy}.
 
Bangladesh makes this gap especially visible. Mobile and digital-platform use is rising among South Asian women, shaped by gender, affordability, and household conditions \cite{gsma2025mobilegendergap, sambasivan2018privacy, ibtasam2019cousin}, and can let mothers encounter health information when formal services fall short \cite{kaur2019engagement, younas2020patriarchy}. But access to information does not guarantee the authority to act on it \cite{sultana2018design, mustafa2020patriarchy}. Formal mental-health provision in Bangladesh is minimal \cite{ bdmentalhealthgap2025}, and postpartum distress ties closely to family conflict, intimate partner violence, weak social support, and material hardship \cite{gausia2009antenatal, valdes2021socioeconomic, azad2019prevalence}, with community studies in Dhaka reporting rates from roughly 24\% at six months to 39\% in the first postpartum year, and stigma keeping women from professional help \cite{hasan2023stigma, chowdhury2024ancient}. HCI4D research adds that women's access to technology and healthcare is shaped by household relationships: mobility restrictions, male approval, shared or intermediated device use, and norms about what women may discuss \cite{sultana2018design, ahmed2019everyone, mustafa2021religion}. So the person who learns about postpartum distress may not have the authority to act on it \cite{alam2025effects}. Bangladesh is therefore a strong context for studying how mothers come to understand postpartum distress, and how that understanding is negotiated within the family.
 
To study this process, we conducted a five-month interview study with 20 mothers in urban, urban-slum, and rural Bangladesh, all within two years of childbirth, asking three questions. \textbf{RQ1:} How do mothers recognize and make sense of postpartum emotional distress, and what part do digital platforms play in that recognition? \textbf{RQ2:} When a mother recognizes her own distress, does that recognition travel to her family, and if not, why not? \textbf{RQ3:} What do the answers to RQ1 and RQ2 imply for the design of postpartum mental-health technologies in family-centered, low-resource settings?
 
We find that recognition is not one event but two. Mothers frequently recognized their own distress only after incidental exposure to postpartum content on Facebook and YouTube that they had not gone looking for, then used these platforms to privately interpret their experience against other mothers' accounts, mostly without disclosing their own. Family recognition was a separate, often unmet, achievement: many mothers could describe their distress clearly and still find that husbands, parents, and in-laws treated it as ordinary. Knowing what was wrong did not translate into the authority to act on that knowledge, and support, when it came, had to be preceded by validation rather than offered on its own. We call this the \textbf{two-level recognition gap}: self-recognition, aided by incidental digital exposure and private sensemaking, does not automatically produce relational recognition, and it is in that gap that distress most often persisted.  
 
This paper makes three contributions to HCI research on digital mental health, feminist HCI, and HCI4D. \textbf{First}, we show how self-recognition of postpartum distress can emerge through incidental exposure to algorithmically recommended content, and how mothers then use digital platforms for low-visibility, private sensemaking rather than public disclosure. \textbf{Second}, we identify the two-level recognition gap and trace it to household epistemic authority: whose knowledge a family treats as legitimate, not simply what a mother knows. \textbf{Third}, we argue that postpartum technologies should support mother-controlled ways of extending recognition to partners, families, and community health workers, rather than treating recognition and disclosure as the mother's task alone.

\section{Related Work} \label{sec:literature-review}

We situate our work across three bodies of literature that together illuminate how postpartum mental health is recognized, supported, and shaped by social context. First, HCI research has explored detecting maternal distress, supporting mothers' digital sensemaking, and designing digital interventions for maternal mental health. Second, research in HCI and CSCW has examined gendered care, family relationships, and household authority, showing how health-related technology use and decision-making extend beyond the individual user. Third, research in South Asian contexts has examined relational approaches to perinatal mental health and digital maternal-health support, highlighting the roles of family, community, and local care structures. Reviewing these strands together allows us to examine what remains less understood about the transition from individual recognition of postpartum distress to recognition and support within the family.

\subsection{HCI Research on Maternal and Postpartum Mental Health}
\label{sec:hci-research-maternal-postpartum-mental-health}

HCI research on maternal and postpartum mental health spans computational detection, digital sensemaking and support, and interventions for screening and care. However, this work remains largely mother-centered and is predominantly situated in Western, urban, and English-language contexts that may differ from South Asian settings.

\subsubsection{Computational Detection of Maternal Mental Health}
\label{sec:computational-detection-maternal-mental-health}

A substantial body of HCI research has explored whether maternal mental-health changes can be detected from digital traces. De Choudhury and colleagues first demonstrated that childbirth is associated with measurable changes in mothers' social-media behavior and language and that these signals can predict postpartum depression (PPD) \cite{dechoudhury2013childbirth,dechoudhury2013predicting,dechoudhury2014characterizing}, building on broader work on computationally inferring depression from social media \cite{dechoudhury2013depression,dechoudhury2016suicidal}. Subsequent studies have used linguistic and visual signals across online platforms to identify mental-health patterns \cite{coppersmith2015clpsych,feuston2018coded,sharma2018mental}. Other systems use passive sensing and wearables to identify maternal stress and mental-health changes \cite{hurwitz2025wear,king2019microstress}, while research in low-resource settings highlights the importance of family involvement, device access, and cultural acceptability alongside technical feasibility \cite{kohrt2019procedures,maharjan2021passive,poudyal2019wearable}. Machine-learning approaches have also been applied to electronic health records for PPD risk prediction \cite{amit2021estimation,zhong2022application}, while HCI research shows that engagement with screening itself is a design concern \cite{doherty2019engagement}.

Alongside these detection-oriented approaches, critical HCI research has questioned the consequences of computational recognition, including the extraction of maternal narratives \cite{antoniak2024nlp}, algorithmic harms in reproductive health \cite{fledderjohann2024algorithmic}, and emotion inference without adequate consent or control \cite{andalibi2020human,roemmich2021data}. Research on fertility applications further shows that simply framing a feature as ``AI-powered'' can influence users' trust \cite{costafigueiredo2024powered}.

Overall, this literature establishes how systems can detect maternal distress, but pays less attention to how mothers themselves recognize and interpret it.

\subsubsection{Digital Sensemaking, Disclosure, and Support-Seeking}
\label{sec:digital-sensemaking-disclosure-support-seeking}

A complementary literature examines how mothers use digital spaces to seek information, make sense of distress, disclose experiences, and obtain support. Analysis of PPD discussions across Reddit, What to Expect, and BabyCenter documents how mothers discuss symptoms, stressors, coping, and social support \cite{progga2023understanding}, while related studies highlight storytelling, peer interaction, and concerns about professional help, anonymity, and misinformation \cite{progga2023therapy,progga2025womens}. Other work examines online health-information seeking and peer support during pregnancy and after pregnancy loss \cite{gui2017investigating,kresnye2020expect,newhouse2016bump2bump}.

Disclosure itself is shaped by platform and social context. Pseudonymous and throwaway accounts can facilitate sensitive parenting disclosures \cite{ammari2019throwaway}, while studies across Reddit, Facebook, and Twitter show that parents adapt self-presentation to different audiences \cite{ammari2015managing,ammari2018pseudonymous,kumar2015babybook}. Research on stigmatized reproductive experiences further describes indirect disclosure or ``testing the waters'' \cite{andalibi2018announcing,andalibi2018testing} and the decision not to disclose despite potential benefits \cite{andalibi2020disclosure}, alongside work on emotional validation and responses to sensitive disclosures \cite{andalibi2018responding,andalibi2021sensemaking}. Thus, privacy, anticipated judgment, and audience are central to how people use digital spaces for sensitive health concerns.

Research also points to inequalities in digital mental-health experiences, including differences in access and satisfaction among perinatal Black women \cite{oguamanam2023intersectional}. In South Asian contexts, Facebook and WhatsApp groups have provided spaces for women to exchange maternal-health information and discuss otherwise difficult experiences \cite{kaur2019engagement,younas2020patriarchy}.

This literature shows how digital spaces support maternal sensemaking and disclosure, but says less about how knowledge or validation gained online translates into support within offline family relationships.

\subsubsection{Designing Digital Interventions for Maternal Mental Health}
\label{sec:designing-digital-interventions-maternal-mental-health}

A third body of research designs digital systems for screening, psychoeducation, self-management, and psychological support. Barry et al. examined self-reporting depressive symptoms during pregnancy and highlighted ethical tensions in mHealth design \cite{barry2017mhealth}, while BrightSelf explored the design of self-report tools and the sharing of wellbeing data between mothers and healthcare systems \cite{doherty2018brightself,doherty2020personal}. Internet-delivered interventions such as MumMoodBooster, Netmums, and Sunnyside have demonstrated the potential of digital psychological support for perinatal depression \cite{duffecy2019sunnyside,milgrom2016internet,omahen2013internet}.

Mobile technologies have also extended maternal-health information through phone-based support and community-led education in low-resource settings \cite{batool2017maternal,kumar2015mobile}. More recent work explores conversational agents for perinatal mental health \cite{mcalister2025chatbot,ramachandran2023woebot,suharwardy2020effect} and culturally situated reproductive-health support \cite{deva2025family,rahman2021adolescentbot,yadav2019feedpal}, alongside self-tracking technologies for parenting stress and pregnancy-related health management \cite{jo2020understanding,lu2024unpacking}. Although most systems remain individually oriented, emerging work considers collaborative care and partners or fathers \cite{haldar2022collaboration,hamil2021development}.

Overall, digital interventions remain largely mother-centered, with limited integration of partners or family members into the same mental-health process.

\subsection{Gendered Care, Family Authority, and Women's Agency in HCI}
\label{sec:gendered-care-family-authority}

While maternal mental-health HCI largely centers the individual mother, HCI and CSCW research shows that \textbf{health, care, and technology use are often distributed across families}. Family-centered health studies examine how caregivers mediate between patients, clinicians, and household members \cite{bhat2020sociocultural,bhat2023halfdoctors}, including in mental health \cite{yamashita2013understanding} and pediatric care \cite{nikkhah2022family}.

Feminist HCI foregrounds \textbf{agency, voice, and often-unrecognized care labor}, challenging narrow assumptions in women's health technologies \cite{bardzell2010feminist,bardzell2011towards,keyes2020reimagining} and advocating more inclusive approaches to women's wellbeing \cite{almeida2016intimate,almeida2020woman,kumar2020taking}. CSCW research similarly shows how essential care and articulation work can remain invisible within technical systems \cite{star1999layers,suchman1995making}, providing a useful lens for understanding the physical, emotional, and organizational labor surrounding postpartum care.

HCI4D research further demonstrates that \textbf{household authority shapes women's access to and use of technology}. Studies in Bangladesh and Pakistan document how mobility restrictions, male approval, and household power influence women's technology use and health decisions \cite{mustafa2020patriarchy,sultana2018design}, while cultural norms shape what reproductive concerns women can discuss \cite{mustafa2021religion}. Intermediated and shared-phone use further shows that access does not necessarily imply autonomous or private use \cite{sambasivan2010intermediated,sambasivan2018privacy}.

Taken together, this literature shows that care, technology use, and health decisions are shaped by family relationships and gendered authority. Yet, this literature rarely examines postpartum mental-health recognition---how mothers' understanding of distress is negotiated within the household, and how family authority and caregiving expectations shape whether it leads to support.

\subsection{Postpartum Mental Health and Digital Support in South Asian Contexts}
\label{sec:postpartum-south-asia}

The relational dimensions identified above are particularly important in South Asian contexts, where research on perinatal mental health has consistently linked maternal wellbeing with marital relationships, family support, household conditions, and gendered social structures \cite{azad2019prevalence,valdes2021socioeconomic}. At the same time, digital maternal-health programs have expanded access to information and support across the region. Reviewing these two bodies of work helps situate the recognition-to-support process within the context in which our study is conducted.

\subsubsection{Relational Approaches to Postpartum Mental Health}
\label{sec:relational-approaches-postpartum}

Research across South Asia links postpartum depression to relational and household conditions, including marital conflict, intimate partner violence, limited social support, and household resources \cite{azad2019prevalence,valdes2021socioeconomic}. This relational framing is also reflected in interventions. The Thinking Healthy Programme (THP) in rural Pakistan involved husbands and families alongside task-shifted psychological care \cite{rahman2008cognitive}, with subsequent trials and reviews supporting community- and family-inclusive approaches across Pakistan, India, and other LMICs \cite{chowdhary2014content,fuhr2019delivering,sikander2019delivering}. Family- and couple-based interventions similarly report benefits from greater family involvement \cite{cluxtonkeller2018clinical,familycentered2023overview}. However, family involvement typically begins after professional identification of distress, leaving less understood how mothers themselves recognize distress and communicate it within the household.

\subsubsection{Digital Maternal Mental Health and Family Contexts}
\label{sec:digital-maternal-family-contexts}

Digital maternal-health initiatives such as Aponjon and Kilkari have expanded access to maternal-health information across South Asia \cite{almamun2019mhealth,chamberlain2021assessing}, alongside mobile and community-based approaches in India, Pakistan, and other resource-constrained settings \cite{aldahdah2015mobile,batool2017maternal,kumar2015projecting}. HCI and ICTD research shows that women's use of these technologies is often shaped by family members, community workers, and household circumstances \cite{batool2017maternal,kumar2015mobile}, with recent work also exploring culturally situated conversational technologies \cite{deva2025family,rahman2021adolescentbot}. Yet postpartum mental health remains comparatively underrepresented. Kamarudin et al. found that mothers with PPD symptoms sought psychoeducation and support for involving partners \cite{kamarudin2024exploring}, while research in Bangladesh highlights stigma, gender norms, and household-mediated technology use as barriers to mental-health support \cite{chowdhury2024ancient,hasan2023stigma,sultana2018design,telemental2025bangladesh}.

Overall, South Asian research highlights the importance of family involvement in perinatal mental-health care. It also shows the growing role of digital technologies in maternal-health support. However, these strands rarely converge around how mothers recognize postpartum distress and seek family support. The relationship between \textbf{individual recognition, family recognition, and support} therefore remains underexplored.

\section{Methodology} \label{sec:methodology}

We explored how Bangladeshi mothers experience and make sense of postpartum emotional distress, and how family, social, and digital contexts shape these experiences, through in-depth semi-structured interviews with 20 mothers aged 18--30, all of whom had given birth within the previous two years. We use \textit{postpartum experience} to mean a retrospectively recalled experience following childbirth, not an ongoing clinical episode; the 24-month upper bound reflects our interest in the broader transition into motherhood, not only the first months after birth. Data were collected over five months (January--May 2026). Participants represented diverse educational, employment, and residential backgrounds and were recruited based on availability and willingness to participate; \autoref{tab:participant-demographics} summarizes their demographics, postpartum experiences, and technology access.

\subsection{Participant Recruitment}
\label{sec:participant-recruitment}

We initially planned to recruit mothers from government hospitals, particularly pediatric wards, but overcrowding and limited privacy at Dhaka Medical College Hospital and a government hospital in Rangpur Division made meaningful conversation difficult; we combined online and offline approaches instead.

\begin{table*}[t]
\centering
\caption{Participant Demographics}
\label{tab:participant-demographics}
\begin{tabular}{lll}
\hline
\textbf{Participant Characteristics} & \textbf{Category} & \textbf{n} \\
\hline
\textbf{Age} & 18--21 years & 7 \\
& 22--25 years & 8 \\
& 26--30 years & 5 \\
\textbf{Child Age} & 3--6 months & 3 \\
& 7--12 months & 6 \\
& 13--18 months & 7 \\
& 19--24 months & 4 \\
\textbf{Education} & Secondary school & 5 \\
& Higher secondary school & 6 \\
& Honors degree or equivalent & 9 \\
\textbf{Employment} & Employed & 3 \\
& Homemaker & 17 \\
\textbf{Residential Setting} & Urban & 10 \\
& Urban slum & 4 \\
& Rural & 6 \\
\textbf{Smartphone Access} & Personal smartphone & 18 \\
& Shared smartphone & 2 \\
\textbf{Internet Access} & Wi-Fi or Mobile Data & 20 \\
\textbf{Postpartum Experiences} & Early/unexpected emotional changes and postpartum distress & 12 \\
& Postpartum depression & 8 \\
& \quad -- Self-diagnosed & 5 \\
& \quad -- Clinically diagnosed & 3 \\
\textbf{Total} & \textbf{Participants} & \textbf{20} \\
\hline
\end{tabular}
\end{table*}

For online recruitment, we shared study invitations in Facebook groups for childcare and women's communities and through the researchers' personal profiles, and contacted mothers who had interacted with postpartum mental-health content online. Invitations described the study and emphasized privacy and confidentiality. Because discussing postpartum mental health was often difficult, given concerns about privacy, judgment, and whether an experience felt legitimate enough to discuss, we emphasized trust-building and confidentiality, which helped some mothers discuss experiences they had previously kept private. This approach reached predominantly urban, well-resourced mothers, several already visible to us through prior online engagement with postpartum content, a limitation we return to in \S\ref{sec:limitations-and-future-work}. To broaden diversity, we separately recruited from an urban slum and a rural community in Chuadanga District through offline contacts: two female university staff members living in the slum, and the first author's familiarity with the rural community.

In total, 32 mothers initially expressed interest; some could not proceed because of childcare and household responsibilities, and a few declined because recalling difficult experiences might cause further distress. Twenty mothers who were willing and comfortable completed the study.

\subsection{Interviews}
\label{sec:interviews-observations}

We conducted semi-structured interviews with all 20 mothers, in person (6) or online (14) via WhatsApp or Messenger, mostly as voice calls, with three preferring text-based conversations over several days. Interviews lasted 40--80 minutes and moved gradually from physical health and daily routines to workload, stress, emotional change, and support, with language adapted to feel less formal in rural and low-income communities. The interview guide was developed and refined with feedback from one of our authors, a physician with prior research experience in women’s healthcare access in Bangladesh.

Recruitment itself revealed contextual challenges that shaped interviews: in the urban slum, families typically lived in single-room homes with shared space and curious neighbors nearby, so we made multiple visits before speaking with mothers alone; in the rural setting, some mothers preferred online communication for the same reason. Privacy was therefore a practical condition, not only an ethical one, shaping whether mothers could discuss postpartum distress at all.

Before each interview, we explained the study and recording procedure and obtained written consent (printed forms in person, Google Forms online), followed by verbal consent before recording. Interviews were conducted in Bengali, and participants were encouraged to describe experiences in their own words rather than label or justify them; we paused or ended an interview whenever participation became uncomfortable.

\subsection{Data Analysis}
\label{sec:data-analysis}

The interviews were transcribed into English by the authors, all of whom are native Bengali speakers.
We then read through transcripts iteratively, removing segments not relevant to our research questions. Analysis followed an inductive, team-based thematic analysis \cite{braun2006using,strauss1990basics}: two researchers coded each transcript independently and without a predetermined codebook, so codes could emerge from participants' own accounts; we treated this independent coding as a basis for analytic discussion rather than a measure of inter-rater reliability. Initial codes included \textit{"unexpected emotional changes"}, \textit{"identity loss"}, \textit{"caregiving burden"}, \textit{"emotional validation"}, and \textit{"family normalization"}, among others, which we grouped into patterns and consolidated into the themes reported in Section~4 across several rounds of team discussion.
Our physician author also provided contextual feedback to inform the interpretation of emerging findings.

\subsection{Ethical Considerations and Positionality}
\label{sec:ethical-considerations-positionality}

Our protocol was reviewed and approved by our institutional research ethics board. We maintained a respectful, non-judgmental environment given the sensitivity and stigma of postpartum experience, informed participants of their right to skip questions or withdraw at any time, and avoided clinical judgments, letting participants describe experiences in their own terms. As researchers affiliated with a University, we recognized that our educational and institutional position could shape both participants' responses and our own interpretations. Being seen as educated, urban-affiliated, and Bangladeshi likely built trust in some conversations and made others more guarded, particularly when mothers discussed husbands or in-laws whose views differed from ours; we return to this in \S\ref{sec:limitations-and-future-work}. We adopted a reflexive approach throughout, and anonymized and generalized sensitive details to protect participants' privacy and safety.

\section{Findings} \label{sec:findings}

This section presents participants' postpartum experiences in five parts: how distress emerged as mothers made sense of unmet expectations of motherhood; how early distress deepened; family and spousal support, and why emotional validation mattered most; digital platforms, which supported recognition and private sensemaking but had limits; and recovery as relational rather than clinical or digital. Participants are identified by number (P\#) throughout; we note residential setting or education level on first mention where relevant.

\subsection{Early Postpartum Experiences Felt Unfamiliar}
\label{sec:early-postpartum-experiences}

Unfamiliar emotional changes, unmet expectations of motherhood, and a fading sense of self shaped the transition before mothers made sense of what was happening.

\subsubsection{Unexpected Emotional Changes}
\label{sec:unexpected-emotional-changes}

For most participants, the first weeks after childbirth brought emotional changes that felt unfamiliar and hard to explain: persistent crying, low mood, irritability, withdrawal, numbness, and a loss of interest in everyday life. The common thread was not just distress, but confusion about why they no longer felt like themselves.
One participant described this shift:

\begin{quote}
\textit{``I felt extremely bad, was constantly sad, and kept crying. I wondered why this had happened to me. Everything felt uncomfortable. It felt like something had happened to my body that I couldn't accept. I was always unhappy, nothing felt good, I didn't feel like doing anything, and I didn't like being around anyone.''} (P10)
\end{quote}

Where some mothers described overwhelming sadness, others described emotional disconnection instead. P17 recalled a month-long \textit{``fog''}---struggling to concentrate, respond to others, or stay connected to daily life---a temporary break, not simply low mood.
Intense, unexplained crying recurred across accounts, even when family actively helped with childcare. P02 explained:

\begin{quote}
\textit{``For the first four to five weeks, I cried almost every day without any reason. I was staying at my mother's house and was well cared for, yet I kept crying. If the baby cried, I cried. If the baby smiled, I cried. Every emotion felt unusually intense.''} (P02)
\end{quote}

For several mothers, distress showed up as volatility rather than continuous sadness: ordinary household events or infant behavior became hard to manage, producing rapid swings between frustration, anger, and tears. P03 described this period as one where \textit{``anger and crying happened together,''} with both feelings arriving at once.

These changes affected relationships too: rather than seeking comfort, some mothers withdrew. P07 described distancing herself from her husband and mother-in-law unless necessary; P10 recalled not leaving home or speaking to anyone outside her household for forty days, repeatedly asking herself \textit{``what had happened to me?''}

\subsubsection{Unmet Expectations of Motherhood}
\label{sec:unmet-expectations-motherhood}

Distress also came from a mismatch between the motherhood mothers had imagined and the one they experienced: most expected immediate happiness and an overwhelming sense of love for their newborn, a gap five participants described in the days right after childbirth. When the expected feelings did not appear, their absence raised doubts about their \textbf{competence as mothers}.

These expectations were often set by idealized portrayals of motherhood, such as childbirth videos and online stories, that shaped what P16 expected her own first moment to feel like. Her experience differed:

\begin{quote}
\textit{``During pregnancy, I watched many childbirth videos on Facebook. Whenever the baby was placed on the mother's chest, the mother would burst into tears with happiness. I even became emotional imagining that moment. But when my own baby was born, I don't know why---I never even looked at my baby on my own. I didn't feel any affection or interest toward her. For the first two or three days, I felt nothing.''} (P16)
\end{quote}

P17 described a similar gap, rooted in a religious framing of the child as a blessing:

\begin{quote}
\textit{``This baby was the reward for everything I had gone through, such a blessing from Allah. How could I feel no emotional pull toward my own baby? Is there something wrong with me?''} (P17)
\end{quote}

Mothers described the bond as developing gradually, but its later arrival did not resolve why they had felt distant at first. With no point of reference---rarely shared, never heard from another woman---the gap could read as evidence they lacked what a \textit{``good mother''} should feel, turning confusion into self-doubt and guilt. P10's account shows this interpretation could outlast the bond itself: \textit{``I never dared to ask anyone. Now when I look at my child, I still feel guilty about those first few days.''}

\subsubsection{Disrupted Identity and Social Visibility}
\label{sec:disrupted-identity-social-visibility}

Motherhood also shifted participants' sense of self: almost every participant described caregiving crowding out personal interests, leisure, friendships, and self-care---not simply becoming busy, but feeling as though their previous identity had disappeared.

P10 described her day as organized around caregiving:

\begin{quote}
\textit{``It was always about the baby---feeding, dressing, cleaning, putting the baby to sleep---all day and all night. It felt like I no longer had a life of my own. I wondered whether my life would always be like this. Would I never be able to go out, chat with people, watch movies, or even take care of myself again?''} (P10)
\end{quote}

At least 14 mothers described struggling to recognize themselves after becoming mothers. P01 reflected that her life felt \textit{``turned upside down''} and that she could \textit{``no longer recognize''} herself.

Family attention also shifted almost entirely to the newborn, leaving mothers' own needs overlooked compared to the care they received during pregnancy. P15 explained:

\begin{quote}
\textit{``When people visit, they rarely ask about me anymore. During pregnancy, everyone was so caring and looked after me. But now, even my husband and my parents rarely ask how I am. They only ask about the baby. It makes me feel ignored.''} (P15)
\end{quote}

P07 described a related loss of social inclusion: \textit{``Nobody checked on me or invited me anywhere because I had a baby. It made me feel very hurt''}.
Mothers did not follow the same emotional trajectory: some recovered within days or weeks, others developed more severe, prolonged distress---in some cases postpartum depression---as caregiving demands, family expectations, identity loss, and physical recovery accumulated over months, and some developed significant distress later despite no early difficulty.

\subsection{How Distress Deepened}
\label{sec:how-distress-deepened}

Distress accumulated rather than staying constant, as stressors intensified after childbirth in different combinations across mothers. Three recurring factors shaped this escalation: relentless caregiving that left little room for recovery, a growing sense that their own lives were on hold, and family hierarchies that valued obedience over mothers' knowledge and judgment.

\subsubsection{Caregiving Without Recovery Time}
\label{sec:caregiving-without-recovery-time}

Caring for a newborn was not, on its own, overwhelming; distress developed when caregiving became relentless, through three conditions: continuous caregiving with inadequate rest, limited practical support despite growing responsibilities, and the expectation to perform complex tasks with little guidance.

\textbf{Continuous caregiving with inadequate rest.} Sleep deprivation was often the first step: mothers described spending entire nights attending to infants who woke frequently to feed, with little chance of uninterrupted sleep, until physical exhaustion gave way to emotional exhaustion.

\begin{quote}
\textit{``My baby hardly slept at night. I had to breastfeed every little while, so I could not sleep the whole night either. This went on day after day. Gradually I started feeling completely helpless, as if I could not continue anymore. Sometimes I even felt it would be better if I left everything.''} (P03)
\end{quote}

For some, caregiving became so continuous that basic physical needs were neglected:

\begin{quote}
\textit{``I couldn't rest because my baby never slept. I had to spend all my time taking care of the baby, day and night. I couldn't sleep, I couldn't even eat properly. My weight dropped from 68 kilograms to 55 kilograms because I simply couldn't manage to eat.''} (P06)
\end{quote}

\textbf{Limited practical support despite growing responsibilities.} In the first weeks, mothers often received temporary help from family, but as they began to recover, this diminished, leaving them as the primary---and in nuclear households, often the sole---caregiver, expected to resume full household responsibilities with little redistribution of labor. P13 felt well supported during two months at her mother's home, but after returning to caregiving and household duties, exhaustion accumulated until, as she put it, \textit{``I stopped wanting to talk to anyone or do anything.''} A similar pattern appeared among mothers with little support from the start: P04, who had raised her ten-month-old largely alone, described caregiving as \textit{``extremely challenging from the beginning,''} a burden that persisted well beyond the newborn period.

Husbands rarely shared childcare or household work. A few participants, mostly urban and more educated, described husbands \textit{``helping''} when home, but this was the exception. One participant captured the inequity: \textit{``Children and families do not belong to a mother alone, so why is the burden of these responsibilities placed solely on her?''}(P18)

\textbf{Expectation to perform complex caregiving tasks with little guidance.} Breastfeeding difficulties were another recurring source of distress: Five mothers described struggling to breastfeed sufficiently. Despite different approaches, relatives' advice, and doctors' consultations, some remained unable to breastfeed sufficiently, affecting their emotional wellbeing.

\begin{quote}
\textit{``My baby has never been able to receive breast milk completely, so I have to give formula milk. I tried everything people suggested. I consulted many doctors, but nothing really helped. Gradually I became very depressed because of this.''} (P11)
\end{quote}

For P20, six months of unresolved breastfeeding difficulties shifted from a feeding concern to a question of competence: she still felt \textit{``like a complete failure as a mother.''}

Three mothers from rural areas linked their breastfeeding difficulties to a lack of practical guidance: many had never been shown how to position the baby, achieve a proper latch, or tell whether breastfeeding was working. One described becoming \textit{``almost frantic with worry''} before teaching herself the correct technique through YouTube. On whether hospital staff had shown her before discharge, she said, \textit{``No. They didn't explain any of those things.''} (P17) Breastfeeding difficulties, then, exposed a gap between the caregiving skills mothers were expected to have and the support they received.

For a smaller group, an infant's ongoing medical complications intensified caregiving demands: three participants cared for medically vulnerable infants through successive health crises. P01, whose baby was born at 33 weeks and later developed pneumonia, described:

\begin{quote}
\textit{``I was still recovering myself, my baby was seriously ill, my partner was away, and whenever something went wrong he blamed me as the mother. Looking back, I don't even know when I gradually fell into depression. The hardest period continued for almost two years.''} (P01)
\end{quote}

When infant health stayed medically complex, and mothers were held responsible for outcomes beyond their control, chances to recover were repeatedly postponed.
\subsubsection{Putting Life on Hold}
\label{sec:putting-life-on-hold}

Beyond caregiving demands, mothers described distress as motherhood reshaped their lives. Loss of autonomy, interrupted education and careers, financial dependence, and limited space for personal needs made some feel that their identities beyond motherhood were fading. This pattern appeared across urban and rural settings, but was less evident in comparatively well-educated and supportive families, where mothers’ personal and professional aspirations were more often accommodated.
For P01, this loss of autonomy took the form of family-imposed restrictions, justified as being for the baby's wellbeing but experienced as a loss of self:

\begin{quote}
\textit{``These problems started after having the baby, when I suddenly went from being free to feeling confined. I could do everything before, but afterwards I felt deprived of everything. There were so many restrictions on what I could eat. I couldn't go anywhere or spend time the way I wanted. More than anything, I felt the absence of my own existence---as if I was no longer a person beyond being a mother. That is where the anger and frustration came from.''} (P01)
\end{quote}

Relatedly, mothers' own needs became hard to justify: an implicit expectation that a ``good mother'' should be selfless made prioritizing herself read as selfish rather than necessary. P09, questioned for wanting to stay fit at twenty-seven, married, and with children, felt as though \textit{``everyone wants me to disappear,''} framing her own health as essential to caregiving instead: \textit{``if I'm not well, how can I take care of my children?''}

For 7 participants, this extended to interrupted education or work. P10, who sat her Higher Secondary exams while pregnant, said: \textit{``I wanted to continue studying after the baby was born and at least complete my honours degree. But raising my baby left me no opportunity. Now it hurts when I see my friends continuing their education while I remain stuck because I never had the chance.''} For P02, her wish to resume studying was dismissed outright by her in-laws, who asked, \textit{``What is the point of studying again after having a baby?''} Similarly, one abandoned her academic career midway through her thesis because \textit{``nobody supports those goals anymore''} (P09); another resigned from her job at her husband's and in-laws' insistence.

These interruptions usually produced financial dependence: participants felt reluctant to voice their needs, like a burden on those supporting them.

\begin{quote}
\textit{``If I asked for anything, they would immediately remind me, 'We're already spending so much on the baby, and on top of that you have all these needs too!' Eventually, I stopped asking for anything.''} (P08)
\end{quote}

\textbf{Rebuilding, with support.} A smaller group (4 of 20), mostly well-educated and urban, reclaimed parts of their previous lives---returning to work, restarting a business, resuming professional goals---linked not to time, but to sustained family support. P11 explained:

\begin{quote}
\textit{``Now that my baby is a little older, and my husband can look after the baby when he is at home, I finally get some time to study and prepare for job opportunities.''} (P11)
\end{quote}
\subsubsection{Knowledge Without Authority}
\label{sec:knowledge-without-authority}

Family relationships shaped mothers' wellbeing not simply through support, but through authority, decision-making, and expectations negotiated after childbirth. Participants described entering motherhood with far greater access to childcare information---through social media, YouTube, online communities, and healthcare content---than previous generations, increasing confidence in evidence-based childcare but creating tension with older family members whose practices were rooted in tradition. Postpartum distress emerged not because mothers lacked knowledge but because they lacked the authority to act on it.
One participant described how childcare disagreements damaged family relationships:

\begin{quote}
\textit{``My in-laws insisted on practices like bathing the baby only on certain lunar dates and using traditional oil massages, while I wanted to follow what I had learned instead. These disagreements gradually created cracks in my relationship with my in-laws.''} (P04)
\end{quote}

Another described how digital knowledge raised her confidence while intensifying conflict with older family members:

\begin{quote}
\textit{``During my pregnancy I watched many doctors' videos on YouTube and Facebook about baby care and nutrition. I knew what was good and what wasn't. But my parents wanted to raise my baby the same way they had raised children years ago. They wouldn't listen to me or even to the doctor's advice. Instead, they argued, 'Didn't we raise children ourselves?' When I tried to explain, I was called disrespectful for talking back.''} (P08)
\end{quote}

For many mothers, conflict extended to the broader question of who was entitled to decide:

\begin{quote}
\textit{``It's my baby. I gave birth to this baby. I stay awake every night breastfeeding. Yet it feels as though I can't make even the smallest decision about my own child.''} (P13)
\end{quote}

Family authority was reinforced through continual criticism of mothers' caregiving competence, judged against what a competent mother should already know, not something learned. As P16 put it, \textit{``This was my first child. I needed time to learn. But the elders never gave me that time. They only kept saying, 'The mother can't do anything.''} For some, criticism shifted from questioning specific practices to undermining their broader competence: P20 was told repeatedly she could not feed, comfort, or care for her baby properly, until she felt like a \textit{``useless mother who knew nothing about my own child,''} despite spending every hour on the infant's care---eroding her confidence and reinforcing dependence on elders' judgment.

When negotiation and correction failed, some mothers began prioritizing autonomy over traditional family support---making childcare decisions independently while still turning to digital resources when needed, rather than rejecting family support outright. As P13 explained:

\begin{quote}
\textit{``I tolerated these judgements for a long time and cried alone many times. Eventually, I decided I would no longer live with either my parents or my in-laws. I moved to where my husband worked so I could raise my baby in my own way. I don't need anyone else's help---I can manage. If I need guidance, I get better advice online.''} (P13)
\end{quote}

Most mothers described a similar trade-off: greater physical demands, since childcare could no longer be shared, but emotional relief, since it restored autonomy over parenting.

\textbf{Where authority was not contested.} Four of 20 reported their childcare decisions were largely respected rather than overridden, which they attributed to their educational background. As P15 put it, \textit{``they know, as I am educated enough I can make the best decision for my child.''} This suggests ``knowledge without authority'' was shaped not only by access to information but by social position: for these four, higher education strengthened the legitimacy of their knowledge in the family's eyes.
\subsection{Negotiating Emotional Support}
\label{sec:negotiating-emotional-support}

Across accounts, emotional validation mattered more than the mere presence of support from family and husbands.

\subsubsection{Seeking Validation Before Support}
\label{sec:seeking-validation-before-support}

Most participants experienced their struggles as normalized within their families: distress was treated as an expected part of motherhood rather than something requiring support. Above all, \textbf{the support they needed was not simply practical help or reassurance that things would improve; they first wanted their emotional experience recognized as real and legitimate}. Several forms of social response made that validation hard to obtain.

\textbf{Fear of judgment.} The most common barrier was expecting that distress would invite criticism rather than understanding. P01 recalled:

\begin{quote}
\textit{``Nobody in my family or relatives understood these issues... My parents would say, `Every mother suffers while raising a child. Why are you being so dramatic?' They compared me with previous generations who raised many children with more hardship, and if I cried or became aggressive, some people even laughed at me.''}
\end{quote}

Such expectations could keep mothers from seeking support even when their distress was severe. P02, despite believing her mother cared for her deeply, could not bring herself to disclose that she needed help: \textit{``I always believed she cared about me the most, yet I still couldn't tell her that I needed her support.''}

\textbf{Protecting personal and family privacy.} Silence was also shaped by concerns about community stigma: in rural communities, mothers anticipated wider criticism if their distress seemed to reject motherhood. P10 noted, \textit{``In villages, people easily criticize mothers for these kinds of things.''} Similar concerns appeared among participants from urban and well-educated families. Although husbands and family members sometimes recognized mothers’ depression and supported professional treatment, they still preferred to keep it within the family because of fears of gossip and reputational harm. P04 explained that although her family supported her treatment, \textit{``they kept the whole thing hidden because they were afraid of what people would say.''} (P04) Privacy, then, was not just confidentiality but protection from what mental illness is taken to mean socially.

This pattern extended to peer relationships too: some mothers disclosed to women they expected to understand, having faced similar struggles, but did not find it. P09 recalled a cousin dismissing her concerns, insisting \textit{``every mother suffers''} and that \textit{``only mothers of the new generation call these things depression.''} Such responses reinforced the expectation that distress should simply be endured.

\textbf{A need for validation, not reassurance.} Not all mothers avoided disclosure out of fear: P07 preferred personal space and feared her family would become overly worried rather than give her the distance she needed. But even mothers who sought support rarely wanted reassurance things would pass; they wanted acknowledgment that the present moment was genuinely hard. P10 said hearing \textit{``Every mother goes through this''} made her feel worse, because she already knew things would change eventually. P09 reflected similarly, \textit{``Forget appreciation --- they couldn't even acknowledge that what I was experiencing was genuinely difficult.''}

Across these interviews, \textbf{validation functioned as a prerequisite for meaningful support}. Mothers were not asking others to solve their distress, but wanted it recognized before advice, reassurance, or intervention followed. As P01 reflected, \textit{``Because we couldn't communicate about it, I never received the help I needed, and it took me much longer to recover.''}

\subsubsection{Husbands as the Preferred Source of Emotional Support}
\label{sec:husbands-preferred-emotional-support}

Participants valued support from parents, siblings, and in-laws, but named their husbands as the person from whom they most wanted emotional understanding. As P05 put it,

\begin{quote}
\textit{``For this time whether anyone else is beside her or not, every woman wants her husband beside her.''}
\end{quote}

Four conditions constrain that support.

\textbf{Physical separation.} Women in Bangladesh commonly spend the early postpartum period at their parental home, or remain apart from husbands working in another city or abroad; 15 of 20 participants described this. These arrangements provided practical help but cost the emotional support participants valued most:

\begin{quote}
\textit{``Being away from my husband during this time created a communication gap. It wasn't possible to share everything over the phone. Besides, if I told him everything, he would only become more worried.''} (P08)
\end{quote}

\textbf{Limited relational groundwork.} Particularly among rural participants, mothers described family pressure to have a child soon after marriage, leaving couples little time to build understanding. P19 put this directly: \textit{``I conceived immediately after marriage, so we never really had the chance to build that understanding. He gave me freedom, but my mental health was simply beyond his understanding.''} Another reflected: \textit{``Wedding, child---everything happened so fast that we couldn't get enough opportunities to know and learn these things (emotional and practical demands of childbirth).''} (P13)

\textbf{Reactive, not proactive, support.} Even willing husbands offered support only once the mother explained her own distress, placing the burden of recognition on already overwhelmed women. P12 summarized this: \textit{``unless I explain everything to him, he cannot understand these problems on his own.''} Several participants described this as an added burden: explaining and educating husbands while struggling with their own well-being. For P07, lacking the capacity for that role, her husband's reactions sometimes intensified rather than relieved her distress.

\textbf{Gender norms and outright denial.} Participants attributed this gap less to indifference than to gender norms limiting men's engagement with emotional wellbeing. As P20 put it, men are often \textit{``taught not to cry or show weakness,''} making it harder to recognize someone else's emotional struggle. For others, limited awareness meant outright denial: P02's husband dismissed the condition as something \textit{``people just make up.''} P01 placed these responses within a broader culture of silence rather than individual husbands' failings:

\begin{quote}
\textit{``Many men in our country have bachelor's or master's degrees, but how many truly understand their wives' struggles? They never learned about these issues within their families. For generations, women have hidden these experiences because of shame and social expectations. So when wives talk about them, many husbands simply say, `We've never seen this happen in our family.' As a result, many women hesitate to share their struggles with their husbands.''} (P01)
\end{quote}

Husbands' support was not fixed: several participants described a shift once husbands recognized postpartum depression or witnessed the severity of distress---suggesting that limited support was awareness, not willingness.

Across this section, support's presence did not automatically translate into feeling supported: family members were often available and husbands willing, but without emotional validation it did not register as support---pushing many mothers toward digital spaces for the acknowledgment they were not finding at home.

\subsection{Navigating Postpartum Experiences Through Digital Technologies}
\label{sec:navigating-postpartum-experiences-digital-technologies}

This section examines how digital platforms became a pathway to discovering postpartum depression, making sense of emotional experiences, and exploring support options.
\subsubsection{Digital Technologies Bridged a Recognition Gap}
\label{sec:digital-technologies-recognition-gap}

Postpartum distress was rarely recognized when symptoms first emerged. Participants often interpreted emotional changes as normal maternal struggles alongside the physical and caregiving demands of early motherhood. Most recognized something was wrong only when symptoms persisted, disrupted daily life, affected their relationship with their baby, or became overwhelming. Prior awareness did not necessarily enable self-recognition: four participants had heard about postpartum depression through education or social media but initially confused their experiences with ordinary maternal stress, while three had never heard the term but sensed something was wrong.

For most, recognition began not through deliberate searching but through incidental exposure to algorithmic recommendations, introducing the concept before mothers went looking for it.

\begin{quote}
\textit{``After my baby was born, I spent most of my time browsing baby-related content on Facebook. Then posts about postpartum depression started appearing automatically. They talked about hormonal changes, strange thoughts, crying for no reason... I began to think that maybe this was exactly what had happened to me after my baby was born.''} (P01)
\end{quote}

Discovering the term often became a turning point, giving participants language for what had previously felt confusing.

\begin{quote}
\textit{``I was watching Facebook Reels when a woman talked about how she protected herself from postpartum depression. That was the first time I had ever heard the term. Out of curiosity, I searched Google, and that day I finally realized what had actually happened to me. Looking back, I think that was the turning point in my whole depression journey.''} (P18)
\end{quote}

This reliance on incidental discovery reflected a gap in routine maternal healthcare, not a preference for online information over clinical care. Participants recalled extensive guidance on pregnancy, childbirth, and physical recovery, but almost none recalled postpartum mental health being raised unprompted: \textit{``nobody at the hospital ever talked about postpartum depression. Only when I learned about it online and asked my doctor did they explain it''} (P12).

Online content often surfaced relatable symptoms without separating postpartum depression from temporary changes such as the baby blues, so some mothers recognized similarities without being sure they had experienced it: \textit{``I don't know whether it was postpartum depression or not, but I experienced something similar for the first few days.''} (P17) Without algorithmic exposure, the same gap could leave distress entirely unrecognized: \textit{``I didn't even know this was a problem. If I didn't know it existed, what would I even have searched for?''} (P08)

Digital technologies filled an informational gap left by routine care, though self-directed discovery meant recognition was uneven, favoring mothers with greater digital access and literacy.
\subsubsection{Digital Technologies Supported Private Sensemaking}
\label{sec:digital-technologies-private-sensemaking}

Once recognized, postpartum depression prompted a shift from passive exposure to active sensemaking: several participants deliberately used digital platforms to understand their symptoms and compare their experience with other mothers', staying private rather than public.

Participants combined resources by need: Google for symptoms and treatment, YouTube for professional explanations, Facebook communities for other mothers' lived experience. They stayed connected to relevant content---reacting to posts or following pages---but rather than joining discussions, most read others' stories to see whether their own experience was shared.

\begin{quote}
\textit{``I used to read those posts regularly because they felt exactly like the things I could never say myself. At least then I didn't feel so alone.''} (P16)
\end{quote}

Reading both professional information and lived experience helped participants reinterpret their emotions as understandable rather than evidence of personal failure. For mothers who felt unable to discuss their struggles with family, digital platforms became a private source of reassurance:

\begin{quote}
\textit{``Since I couldn't get support from my husband or family, the internet became my only option. It really helped me understand my problem and gradually regain my confidence.''} (P19)
\end{quote}

Even so, participants were reluctant to disclose their struggles online, citing privacy: \textit{``I don't like bringing personal matters online... I'm simply not comfortable sharing such personal problems publicly''} (P05). They worried a Facebook post could expose them to relatives' judgment, affect how their family was seen, or draw an unwelcome response from a husband or in-law. Participants perceived risk even in anonymous sharing, since negative comments from strangers could still hurt and deepen distress, regardless of whether others could identify them. The value of these platforms lay not in enabling public help-seeking but in offering a private space to validate emotion and reduce self-blame, while mothers kept full control over whether, when, and with whom to disclose further.

\subsubsection{Mothers Strategically Evaluated Digital Resources}
\label{sec:mothers-strategically-evaluated-digital-resources}

Mothers also evaluated digital resources on several fronts: credibility, language accessibility, ease of use, and, for those who had considered AI tools, a risk of premature self-diagnosis.

\paragraph{Credibility.}
Google and YouTube were generally trusted to point mothers toward healthcare professionals and evidence-based content. Facebook was more ambivalent: valuable for discovery and peer-comparison, but distrusted as medical fact: \textit{``there are many bloggers on Facebook who have no medical background. Some portray postpartum depression as something extremely negative, while others say it's nothing at all''} (P07). P02 put it plainly: \textit{``Google is better for finding information because it usually provides more accurate content. On Facebook everyone leaves different comments, so it's difficult to know what is actually correct.''}

\paragraph{Language access.}
Participants who searched for content found a scarcity of quality Bangla-language material, even where English content was abundant: \textit{``I could hardly find any professional content in Bangla. When I searched on Facebook, there was almost nothing. Even on YouTube, whatever content was available was very brief''} (P12). This compounds the digital-access inequality above: mothers fluent in English reached a larger, more credible information base than those who were not.

\paragraph{AI-specific risk.}
A few participants who had considered tools like ChatGPT worried AI might amplify an unverified self-diagnosis, raising anxiety rather than resolving it. P17 reflected, \textit{``if I only suspected I had a problem and ChatGPT simply confirmed it, I would become even more anxious.''}

\paragraph{Accessibility over sophistication.}
Participants also judged technologies by how easily they became useful, distinguishing platforms less by capability than by how much prior knowledge each demanded. P01, who had not used AI tools herself, contrasted them with Facebook and YouTube:

\begin{quote}
\textit{``I have never used AI or ChatGPT. Most mothers, especially in rural areas, are much more familiar with Facebook and YouTube. With AI, you first have to know what to ask. But on Facebook or YouTube, relevant videos appear automatically and keep being recommended. Even mothers with little education or limited digital skills can still use those platforms.''} (P01)
\end{quote}

Taken together, mothers weighed credibility, language access, emotional risk, and usability at once---a more active process than ``searching online'' suggests.
\subsection{Recovery as Relational, Not Clinical or Digital}
\label{sec:recovery-relational-not-clinical-digital}

Participants rarely described postpartum depression as resolved primarily through the healthcare system, and digital recognition and self-directed coping were not sufficient alone. We report this as what participants needed and experienced, not a general claim about how postpartum care should be sequenced; we return to that distinction below.

\textbf{The Layered Model: Family First, Clinical Support Second.} Participants described a consistent sequence: emotional support from husbands, practical help from family, and opportunities for rest came first, with professional treatment sought only once these proved insufficient. As P11 put it, \textit{``since many of these problems are created within the family, I think the solution should also begin within the family.''} P15 described the same order: family support came first, and \textit{``only if the situation becomes very difficult should professional treatment become necessary.''}

This was less a rejection of healthcare than a belief about where distress originated: in family interactions, caregiving demands, and the home environment rather than an individual medical condition, so family support was a necessary starting point and clinical treatment insufficient while those conditions stayed unchanged.

One account complicates this sequence rather than confirming it. P06, reflecting after a severe depressive episode, concluded that delaying professional care had let her condition worsen: \textit{``if I had sought help earlier, maybe it wouldn't have become this severe.''} Read alongside the family-first pattern, her account suggests participants often recognized the value of timely clinical care only in hindsight. We take this as a caution, not a contradiction: a family-first sequence carried real risk when support was slow to arrive, and should not be read as evidence family support ought to precede professional care.

\textbf{Healthcare evaluated by family engagement, not diagnosis alone.} Participants evaluated clinicians less by diagnostic accuracy than by whether care extended into the family relationships shaping their depression. P03 described a gynecologist consultation as transformative because the physician engaged her husband directly, not only her.

\begin{quote}
\textit{``She took time to explain postpartum depression to my husband as well. I think because she treated the situation so seriously, my husband and family also began to understand it seriously.''} (P03)
\end{quote}

P02, by contrast, sought clinical care only after repeatedly asking her husband, who did not believe in postpartum depression. She was eventually taken to a \textbf{nearby government hospital}, where a gynecologist recognized her experience as postpartum depression and advised rest and self-care, with medication if symptoms worsened. But the consultation left the family context she had described unaddressed:

\begin{quote}
\textit{``My husband was waiting on the other side of the door while I was talking to the doctor.... I was expecting her to call my husband and explain everything to him so that he could at least understand that my problem was real, but she didn't call him. Ultimately, it did nothing for me.''} (P02)
\end{quote}

Clinical recognition felt meaningful only when it translated into family understanding too.

\textbf{The limits of self-directed digital recovery.} Section~4.4 showed digital platforms let mothers recognize and privately validate distress even without family support; this section complicates that picture, since recognition and self-validation alone were often not enough to sustain recovery. Some mothers tried to use digital information as a bridge into family understanding: P20 shared postpartum-depression content from Facebook with her husband to explain her experience, but the attempt was dismissed instead: \textit{``you are going mad by seeing these contents''} (P20).

Where that translation failed, participants attempted recovery largely alone through self-care strategies drawn from social media, struggling to sustain them without family support---14 of 20 described this pattern. This compounded their distress rather than resolving it: having recognized their condition as valid but still lacking the family support they needed, some felt worse for having gained that awareness. One put it simply: \textit{``it was better to not know''} (P02).

Taken together, these accounts suggest digital technologies helped mothers recognize and validate their experiences, but did not substitute for the relational support they considered essential to recovery.

\section{Discussion} \label{sec:discussion}

\begin{figure*}[t]
    \centering
    \includegraphics[width=0.98\textwidth]{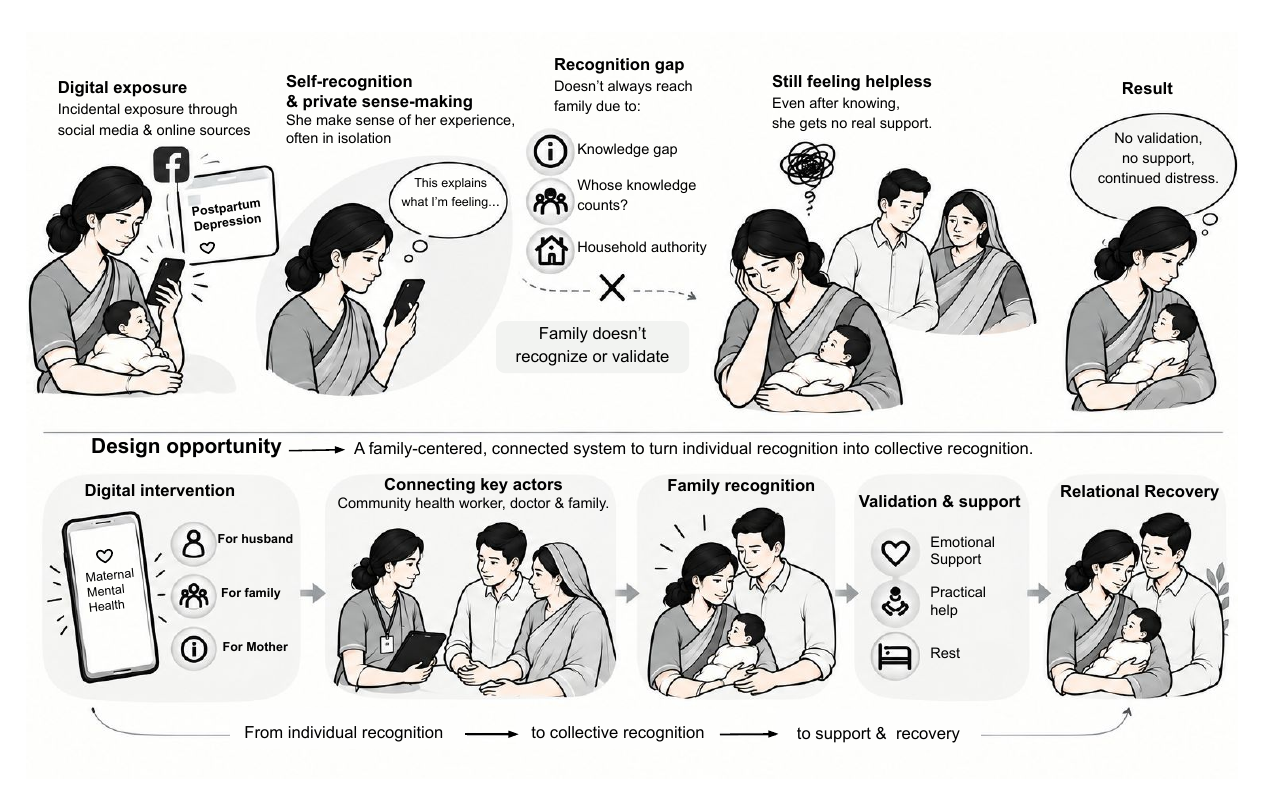}
    \caption{
    From individual recognition to relational recovery. The top pathway shows how incidental digital exposure can support mothers’ self-recognition and private sensemaking, but gaps in knowledge and household authority may prevent this recognition from becoming family recognition, leaving mothers feeling helpless. The bottom pathway presents a design opportunity to connect mothers with husbands, family members, doctors, and community health workers, enabling collective recognition, validation, practical support, and rest toward relational recovery.
    }
   \Description{
    A left-to-right conceptual diagram showing two pathways from postpartum mental-health recognition to recovery. The top pathway shows digital exposure through Facebook, YouTube, and online stories leading to mothers' self-recognition and private sensemaking. A gray barrier labeled ``Recognition Gap'' separates individual recognition from family recognition, representing household authority and whose knowledge is treated as legitimate. Family recognition leads to validation, practical support, and opportunities for rest, resulting in relational recovery. A doctor or gynecologist is shown above the recognition gap as a pathway that can facilitate family engagement. The bottom pathway shows a proposed design opportunity connecting mothers with husbands, family members, doctors, and community health workers to extend individual recognition into collective recognition, validation, and relational recovery.
}
    \label{fig:relational-recovery-model}
\end{figure*}

Our findings show how gendered caregiving, family authority, and limited recognition shape mothers' distress and help-seeking in Bangladesh. Digital technologies supported awareness and private sensemaking, yet this knowledge did not reliably translate into family recognition or support. We use that gap to argue that postpartum technologies should support mothers' privacy while also engaging the family contexts in which their knowledge is interpreted and acted upon.
\autoref{fig:relational-recovery-model} synthesizes this process. It shows how digital platforms and peer experiences can support self-recognition, while the transition from individual recognition to family recognition is mediated by household authority, validation, and relational support. The figure also distinguishes observed pathways in our participants' accounts from potential design pathways involving family, healthcare providers, and community health workers.

\subsection{Individual Recognition Does Not Automatically Become Family Support}

Family members, particularly husbands, were often participants' first source of support, but support was meaningful only when it involved recognition and responsiveness, not assistance alone, consistent with evidence linking partner support to perinatal mental health \cite{pilkington2015partner, jones2013southasian}. Our findings identify two levels of recognition: mothers recognizing their experience exceeds ordinary difficulty, and family members recognizing the distress is real. Digital platforms could facilitate the first: participants encountered PPD-related information through Facebook, YouTube, and Google, consistent with HCI research on postpartum awareness \cite{progga2023understanding, progga2025womens}. But exposure did not always produce immediate self-recognition.

The second level was harder: participants anticipated judgment, and family members normalized the distress rather than recognizing a need for support, extending prior evidence on stigma and help-seeking barriers \cite{dennis2006postpartum, sword2008women} and on under-support of fathers and partners in parenting technologies \cite{matthews2022fathers, bhat2025fatherhood}. A mother could become better informed without gaining the rest, autonomy, or redistributed caregiving recovery requires, and unmet needs made salient by awareness could intensify helplessness. This gap was not universal: a few participants described husbands or mothers-in-law who responded to early distress without digital information first, usually because that relative had prior experience of a difficult postpartum period. It was a minority pattern, suggesting the gap reflects family experience rather than inevitability.

The implication for HCI is not that mothers need more information, but that technologies designed only to inform them leave the family gap untouched. Postpartum technologies could instead create opportunities for partners and family to develop awareness and participate proactively in care; \S\ref{sec:design-implications} outlines what this could look like.

\subsection{Family Recognition Depends on Whose Knowledge Counts, Not Just What Is Known}
\label{sec:epistemic-authority}

Why does mothers' recognition not transfer to others? We find it useful, though not conclusive, to read our findings alongside Fricker's epistemic injustice: \textit{testimonial injustice}, where an account is given less credibility because of who holds it, and \textit{hermeneutical injustice}, a lack of conceptual resources for understanding an experience \cite{fricker2007epistemic, carel2014epistemic, crichton2017epistemic}. We use these as an interpretive lens rather than a claim that we measured injustice directly. Our interviews document that mothers' experiential knowledge was not always treated as authoritative, and information from doctors or digital sources did not reliably close this gap: the barrier was whose knowledge counted in family decisions, not access to information \cite{mccollum2012hermeneutical}.

Education appeared to shape authority: more educated mothers were often trusted in childcare decisions, consistent with research linking women's education and economic resources to decision-making autonomy in Bangladesh \cite{rahman2014autonomy, raz2025autonomy, shourove2023autonomy}, though this varies \cite{jayachandran2026autonomy, kirkwood2024decision}. Knowledge alone did not guarantee authority: social position and perceived competence shaped whether it was heard, so more information may matter less than whether mothers hold the authority to act on it.

A related pattern appeared in self-understanding: some mothers read emotional disconnection as personal inadequacy rather than a symptom, consistent with research on how ``good motherhood'' norms discourage help-seeking \cite{dennis2006postpartum, sword2008women}. For P17, this took an explicitly \textbf{religious form}: framing a child as a divine blessing left no space to also feel emotionally distant, so disconnection read as a personal failing rather than a recognizable symptom (\S\ref{sec:unmet-expectations-motherhood}). Prior HCI mental-health design has largely been built for Western, secular contexts \cite{linxen2021weird}; where religious meaning-making is how mothers first interpret their experience, recognition tools built around a purely clinical vocabulary may not reach them. That means the challenge was not simply lacking information, but lacking a framework that made experiences understandable and legitimate. This is why mothers also emphasized \textbf{acknowledgment before reassurance}: ``every mother goes through this'' can normalize distress without recognizing it as real.

The gap also has an intergenerational dimension: grandmothers and mothers-in-law often hold influence through established practice \cite{negin2016grandmothers, pradhan2023motherinlaw}, while younger mothers have more access to current information, a tension that extends feminist HCI's concern with whose knowledge is centered \cite{bardzell2010feminist, dignazio2016feminist}. Postpartum technologies should support the family conditions through which knowledge is recognized, consistent with calls to integrate maternal mental-health support into routine community and clinical care \cite{who2022integration, rahman2019capacities}.

\subsection{The Recognition Gap Is Gendered by Design of the Household, Not by Accident}

The preceding subsections show \textbf{how} the gap occurs; this one examines why it is gendered. Women carry a disproportionate share of household and care work in Bangladesh and beyond \cite{afrin2024gender, reichstiebert2023gendered}, and husbands' contributions were often framed as \textit{``help''} rather than shared responsibility, leaving mothers to also explain why their distress warranted attention \cite{ammari2017diy, bhat2025fatherhood}. Feminist HCI's focus on whose needs are centered \cite{bardzell2010feminist} and Bangladesh-focused HCI research on patriarchal family structures \cite{sultana2018design} situate this pattern: digital support cannot be separated from the family structures that determine whether mothers can act on what they know.

This structure may also shape how couples enter parenthood: several described family pressure to have a child soon after marriage, since delay in Bangladesh can raise concerns about female fertility and reputation \cite{gipson2015sociocultural, huda2020understanding}, leaving couples less time to build shared expectations, consistent with research on the transition to parenthood \cite{delicate2018systematic, barclay2012preparing}. We do not read this as evidence that early childbirth causes poor spousal support, only that less preparation time may leave husbands less able to anticipate their partners' needs.

This also sits within gendered norms of privacy around reproductive health, which limit husbands' knowledge and involvement \cite{rahman2018knowledge, henshaw2016partner, insan2022perinatal}. We offer, as an interpretation rather than a causal claim, a possible self-reinforcing cycle: when women's health is managed privately, men have fewer chances to learn what it involves, leaving them less prepared to recognize distress and further reinforcing private management. For HCI, this suggests postpartum support should \textbf{make mothers' often-invisible struggles more visible to partners} before distress becomes acute \cite{grimme2026navigating, yarosh2016parenting}.

\subsection{Digital Platforms Widen Access to Information Without Guaranteeing Its Relevance}

Awareness often emerged without deliberate searching: mothers encountered Postpartum depression-related content while browsing pregnancy or childcare material \cite{tian2008media, tian2009media}, and Facebook and YouTube recommendations sometimes surfaced a condition they did not yet know to search for \cite{lim2022socialmedia, yu2024recommendation}. This incidental exposure is an opportunity, not a guarantee: recommendation systems respond to behavioral signals in ways that are neither deterministic nor uniformly beneficial \cite{metzler2024social, holtz2020engagement}, and exposure does not ensure the information is credible. The HCI challenge is not simply making Postpartum distress visible, but determining what becomes visible and to whom it is relevant.

This reliance on digital discovery is notable: no participant reported receiving postpartum mental-health information from a healthcare provider, consistent with evidence that perinatal mental-health care remains poorly integrated into maternal healthcare in Bangladesh \cite{dutta2022mental, insan2023perinatal}. Digital platforms partially compensated for this gap, unevenly, since benefiting from online health information depends on access, eHealth literacy, and socioeconomic position \cite{neter2012ehealth, khatun2017gender, ahmed2020digital}: what is discoverable is not always accessible to all mothers. This suggests an HCI opportunity to support the discoverability of credible, culturally and linguistically appropriate information within existing routines, alongside strengthening its integration into routine maternal care through brief provider guidance, referral pathways, and community health workers, aligning with Bangladesh’s community-based national mental health policy \cite{bdmentalhealthpolicy2022}. Digital platforms could reinforce these pathways rather than replace them.

\subsection{Silent, Private Use Is Meaningful Participation, Not a Failure to Engage}

Digital platforms also provided a private space for sensemaking, extending research on how postpartum communities support information exchange and reduce isolation \cite{chivers2021support, grimme2026navigating}: mothers benefited without disclosing their own experiences, comparing others' accounts with their own while controlling disclosure.

This preference reflected concerns about being identified. Although anonymity can facilitate disclosure of stigmatized experiences \cite{andalibi2018announcing, moore2017use}, it does not remove disclosure's emotional risk: participants worried about how their experiences might be interpreted even unidentified. Participation should not be equated with posting: silent reading is itself meaningful engagement.

For HCI, this points to graduated, user-controlled participation rather than disclosure by default: a continuum from private reading to anonymous, pseudonymous, or identifiable sharing \cite{ma2016anonymity, xu2025influence}, with active moderation since anonymity alone does not prevent harmful responses \cite{potnis2022information, dwork2024moderation}.

\subsection{Family-Centered Design Cannot Mean Mandatory Family Access}
\label{sec:design-implications}

Much existing postpartum mental-health design assumes an individual user, professional referral pathways, and secular clinical language \cite{linxen2021weird}, assumptions that fit poorly where family authority governs care and religious framing shapes how distress is first understood \cite{mustafa2021religion}. Taken together, our findings also do not support routing mothers' data or disclosures to family members by default: the same household authority structures that made mothers' knowledge easy to dismiss (\S\ref{sec:epistemic-authority}) could make family-visible health data another site of control rather than support. We therefore treat family-centered design as an invitation-based complement to mother-controlled private use, not a replacement for it, and outline \textbf{four principles} that follow from our findings.

First, \textbf{keep private use the default, and make family visibility opt-in and reversible}: since silent, undisclosed use was itself meaningful participation, any feature surfacing a mother's activity to family should require her explicit, revocable consent per recipient.

Second, \textbf{design separate entry points for partners and family rather than a shared account}: husbands and mothers-in-law often lacked a framework for recognizing distress, not just information about one mother, so a family-facing feature could build that awareness without exposing a mother's private disclosures.

Third, \textbf{support choosing what, when, and to whom to disclose}, rather than treating disclosure as all-or-nothing (sharing a symptom checklist without a journal entry, for instance), respecting the distinction participants drew between wanting acknowledgment and wanting to remain unobserved.

Fourth, \textbf{treat family involvement as context-dependent, and design for its absence}: some participants described family responses that dismissed their distress; a system assuming a responsive family member, or nudging disclosure, risks harm in unresponsive or unsafe households. Family-facing features should stay optional additions to a system that works fully alone.

These principles work within existing family and platform dynamics; they cannot by themselves shift the gender norms, patriarchal authority, or digital-access gaps that produced the recognition gap we describe, and closing it fully depends on that wider change too \cite{sultana2018design, islam2021gendered}.

\section{Limitations and Future Work} \label{sec:limitations-and-future-work}
Two limitations concern who our findings speak for. First, part of our sample was recruited online (\S\ref{sec:methodology}), reaching predominantly urban, digitally engaged mothers, some of whom were already visible to us through prior engagement with postpartum content. Our findings on digital technology use are most representative of this group and, reflecting mainly Facebook and YouTube use, may not extend to platforms with different logics. Second, being seen as educated, urban-affiliated researchers likely built trust in some conversations and made others more guarded, especially when mothers discussed husbands or in-laws whose views differed from ours. We also recruited from an urban slum and a rural community through offline contacts, but access relied on intermediaries and the first author's familiarity with her home village, and socioeconomic differences between researchers and participants may still have shaped disclosure despite repeated visits and rapport-building.

Further constraints bound our study. A measles outbreak during the study limited our ability to reach mothers in urban slums and villages and to revisit some participants as intended \cite{who2026measles}. All participants were first-time mothers; we intentionally excluded mothers who felt recalling their postpartum experiences could cause further distress, likely underrepresenting the most severe cases, and our sample was predominantly homemakers. We therefore offer a situated study bounded by these communities and this period, not a generalizable picture of postpartum distress in Bangladesh, and our design directions remain proposals awaiting empirical evaluation, not tested interventions.

Future work should test these design directions directly: whether mother-controlled disclosure tools change how partners and family respond, not only whether mothers adopt them; how husbands and other family members can be engaged in postpartum support and sustained in that role over time; and, longitudinally, how the recognition gap, awareness, and technology use shift as digital literacy grows.

\section{Conclusion} \label{sec:conclusion}
This study examined how mothers in Bangladesh experience postpartum distress, seek recognition, and use digital technologies to make sense of what is happening to them. We show that self-awareness is only half of what recovery needs. Many mothers named their own distress clearly and still returned to a household that treated it as ordinary; that second gap, not the first, is where support most often failed. Digital platforms can help mothers reach the first kind of recognition. They cannot, on their own, produce the second.

Our findings contribute to maternal mental-health HCI, HCI4D, and feminist HCI: postpartum technologies must extend beyond individual awareness toward recognition, communication, and shared responsibility within families, while leaving mothers in control of if, when, and with whom support is shared. Closing the recognition gap means designing technologies that work with the family, not only for the mother: tools that keep her privacy the default while giving her real, controllable ways to extend what she knows to the people whose recognition her recovery depends on. That is the design problem this paper leaves for HCI to take up: not how to reach mothers, but how to help their families finally hear them.

\bibliographystyle{ACM-Reference-Format}
\bibliography{references}

\end{document}